\documentclass[webpdf,modern,large]{oup-authoring-template}

\graphicspath{{Fig/}}

\theoremstyle{thmstyleone}%
\theoremstyle{thmstyletwo}%
\theoremstyle{thmstylethree}%

\usepackage{hyperref}
\usepackage{url}
\usepackage{graphicx}
\usepackage{acronym}  
\newacro{MethodName}[MFD]{Metadata-guided Feature Disentanglement}

\newacro{MI}{Mutual Information}
\newacro{DL}{Deep Learning}
\newacro{CNN}{Convolutional Neural Network}
\newacro{DNN}{Deep Neural Network}
\newacro{ANN}{Artificial Neural Network}
\newacro{AUROC}{Area under the Receiver Operating Characteristic}
\newacro{DRL}{Disentangled Representation Learning}
\newacro{MLP}{Multilayer Perceptron}

\newacro{ENCODE}{Encyclopedia of DNA Elements}
\newacro{eQTL}{Expression Quantitative Trait Loci}
\newacro{GTEx}{Genotype-Tissue Expression}
\newacro{MAF}{Minor Allele Frequency}
\newacro{VEP}{Variant Effect Prediction}
\newacro{OR}{Odds Ratio}

\newacro{XAI}{Explainable Artificial Intelligence}
\newacro{TF}{Transcription Factor}
\usepackage{caption}
\usepackage{subcaption}
\usepackage[inline]{enumitem}

\begin{document}
\newcommand{\E}{\mathbb{E}}
\newcommand{\Ls}{\mathcal{L}}
\newcommand{\R}{\mathbb{R}}
\newcommand{\emp}{\tilde{p}}
\newcommand{\lr}{\alpha}
\newcommand{\reg}{\lambda}
\newcommand{\rect}{\mathrm{rectifier}}
\newcommand{\softmax}{\mathrm{softmax}}
\newcommand{\sigmoid}{\sigma}
\newcommand{\softplus}{\zeta}
\newcommand{\KL}{D_{\mathrm{KL}}}
\newcommand{\Var}{\mathrm{Var}}
\newcommand{\standarderror}{\mathrm{SE}}
\newcommand{\Cov}{\mathrm{Cov}}

\newcommand{\latentSize}{C}
\newcommand{\numMetadataVars}{M}
\newcommand{\numSamples}{N}
\newcommand{\numClasses}{O}
\newcommand{\featuresBio}{\mathbf{s}_{bio}}
\newcommand{\featuresTech}{\mathbf{s}_{tech}}
\newcommand{\featuresFull}{\mathbf{s}_{full}}
\newcommand{\predFn}{\Phi}
\newcommand{\LNorm}{{l}}
\newcommand{\numExperimentsNumeric}{2,106}

\def\PrintMainText{1}
\def\PrintAppendix{1}


\journaltitle{Journal Title Here}
\DOI{DOI HERE}
\copyrightyear{2022}
\pubyear{2019}
\access{Advance Access Publication Date: Day Month Year}
\appnotes{Paper}

\firstpage{1}


\title{\acl{MethodName} for Functional Genomics}

\author[1,$\dagger$]{Alexander Rakowski}
\author[1,2,$\dagger$]{Remo Monti}
\author[2]{Viktoriia Huryn}
\author[3]{Marta Lemanczyk}
\author[2,$\ast$]{Uwe Ohler}
\author[1,4,$\ast$]{Christoph Lippert}

\authormark{Rakowski, Monti et al.}
\address[1]{\orgdiv{Digital Health Machine Learning}, \orgname{Hasso Plattner Institute for Digital Engineering}, \orgaddress{\street{Rudolf-Breitscheid-Str. 187}, \postcode{14482}, \state{Potsdam}, \country{Germany}}}
\address[2]{\orgdiv{Max-Delbrück-Center for Molecular Medicine in the Helmholtz Association}, \orgname{Berlin Institute for Medical Systems Biology}, \orgaddress{\street{Robert-Rössle-Str. 10}, \postcode{13125}, \state{Berlin}, \country{Germany}}}
\address[3]{\orgdiv{Data Analytics and Computational Statistics}, \orgname{Hasso Plattner Institute for Digital Engineering}, \orgaddress{\street{Prof.-Dr.-Helmert-Str. 2-3}, \postcode{14482}, \state{Potsdam}, \country{Germany}}}
\address[4]{\orgdiv{Hasso Plattner Institute for Digital Health at Mount Sinai}, \state{New York}, \country{United States}}

\corresp[$\dagger$]{Both authors contributed equally, and are listed in random order\\}
\corresp[$\ast$]{Corresponding authors: \href{email:christoph.lippert@hpi.de}{christoph.lippert@hpi.de}
\href{email:uwe.ohler@mdc-berlin.de}{uwe.ohler@mdc-berlin.de}}




\abstract{With the development of high-throughput technologies, genomics datasets rapidly grow in size, including functional genomics data.
This has allowed the training of large \ac{DL} models to predict epigenetic readouts, such as protein binding or histone modifications, from genome sequences.
However, large dataset sizes come at a price of data consistency, often aggregating results from a large number of studies, conducted under varying experimental conditions.
While data from large-scale consortia are useful as they allow studying the effects of different biological conditions, they can also contain unwanted biases from confounding experimental factors.
Here, we introduce \ac{MethodName} - an approach that allows disentangling biologically relevant features from potential technical biases.
\ac{MethodName} incorporates target metadata into model training, by conditioning weights of the model output layer on different experimental factors.
It then separates the factors into disjoint groups and enforces independence of the corresponding feature subspaces with an adversarially learned penalty.
We show that the metadata-driven disentanglement approach allows for better model introspection, by connecting latent features to experimental factors, without compromising, or even improving performance in downstream tasks, such as enhancer prediction, or genetic variant discovery.
The code for our implemementation is available at \url{https://github.com/HealthML/MFD}
}
\keywords{functional genomics, epigenetics, disentanglement, representation learning}

\maketitle
\section{Introduction}
Consortia such as \ac{ENCODE}~\citep{encode2020} have accumulated a wealth of high-throughput functional genomics data across a broad range of cell lines, developmental time points, and tissues, for instance measuring chromatin modifications and DNA accessibility. These data have spurred the development of deep neural networks that predict the readouts of these experiments from DNA sequence inputs to better understand the sequence features that govern gene regulation~\citep{zhou2015deepsea,kelley2016basset,avsec2021bpnet}.

The development of \ac{XAI} methods has allowed for assessing the importance of input features for \ac{DL} models' predictions. A commonly used approach to interpret genomic deep learning models comprises post hoc interpretation methods, producing sequence attribution maps (for an overview, see~\citep{novakovsky2023obtaining}). However, these maps have been shown to produce spurious results~\citep{hooker2019benchmark}. Although properties of the learned function and the particularities of the methods themselves have been identified as contributing to noisy attributions, and solutions have been proposed~\citep{majdandzic2023PKoo}, these do not tackle the issue of noise in the training data.

Genomics data are heavily affected by experiment-specific (e.g., selectivity of DNA restriction enzymes) and technology-specific (e.g., adapter choice, amplification method) biases as well as strong batch effects (e.g., laboratories, vendors)~\cite{leek2007capturing}. These biases mask intended signals and affect downstream analyses. Proposed correction methods usually address only specific sets of biases and have not become widely used in practice~\citep{Wang2017Correcting}. Recent work has demonstrated the utility of \ac{XAI} to uncover biases in genomics training data~\citep{ghanbari2020deep}, which indicates that genomics models may heavily rely on biases in addition to genuine biological features to make predictions. It is unclear how strongly this affects downstream applications, such as enhancer sequence or genetic variant effect prediction. Therefore, directly modeling sources of bias and employing inherently interpretable model designs should contribute to overcoming these issues and improving downstream task performance.

\ac{DRL} focuses on separating the generative factors underlying the observable data~\citep{bengio2013representation} by imposing properties on a learned latent data representation space, e.g., conditionally factorizable priors~\citep{locatello2020weakly,khemakhem2020variational}, or imposing invariance to a set of variables~\citep{ganin2016domain,zhao2020training,adeli2021representation,he2021code}. The recently introduced method, Disentangled Relevant Subspace Analysis (DRSA)~\citep{Chormai2022DisentangledEO}, enhances the interpretability of machine learning models by working in conjunction with XAI techniques. DRSA focuses on analyzing relevant subspaces within a model's activation layers rather than solely examining the final predictions. This approach separates and clarifies the contributions of various features to model decisions, enhancing transparency and understanding of complex datasets.

In the context of biomedical applications, \ac{DRL} models demonstrate increased explainability, robustness, and better generalization~\citep{schreiber2020avocado,YANG2022989deepnoise,lotfollahi2023biologically}. Such approaches typically require information on a per-observation level, typically in the form of additional observed variables. Instead, we consider a setting where the auxiliary information is not available per-observation, but we have access to metadata defining relations between different classes of outcomes.


To this end, we propose \acf{MethodName}—a \ac{DNN} DNA sequence model that leverages metadata of the predictions of interest, in our case metadata from \ac{ENCODE} experiments, to separate biological features from technical ones by learning two independent latent subspaces. We train \ac{MethodName} on human genome data to predict peak calls from $\numExperimentsNumeric$ \ac{ENCODE} experimental tracks, and we demonstrate its impact on model interpretability (Section~\ref{ssec:interpretability}) and downstream task performance on independent data (Sections~\ref{ssec:enhancer} and~\ref{ssec:vep}).

\section{\acl{MethodName}}
\newcommand{\MetadataEmbedding}{\psi}
\ac{MethodName} is a \ac{DL} model predicting peak calls 
of multiple tissue-based experiments from DNA sequence data while learning two disentangled feature sub-spaces, corresponding to biological and technical experiment metadata.
It consists of 3 modules: 
\begin{enumerate*}[label=(\roman*)]
\item a \ac{CNN} sequence feature extractor, based on the Basenji2 architecture~\citep{avsec2021effective}
\item a metadata embedding module based on two static hypernetworks~\citep{ha2016hypernetworks} mapping the metadata of each experiment to a set of weights, which are in turn used to compute the corresponding peak prediction from the extracted DNA features (Section~\ref{ssec:model_metadata_embeddings})
\item a regularization penalty, enforcing independence between the two latent sub-spaces of the model (Section~\ref{ssec:model_disentanglement})
\end{enumerate*}.
Model training and data collection are described in Appendix 
sections A and C.
\subsection{Metadata Embeddings}
\label{ssec:model_metadata_embeddings}
\begin{figure*}
\centering
\begin{subfigure}{.5\textwidth}
  \centering
  \includegraphics[trim={0 6.5cm 0 0},clip,width=1\linewidth]{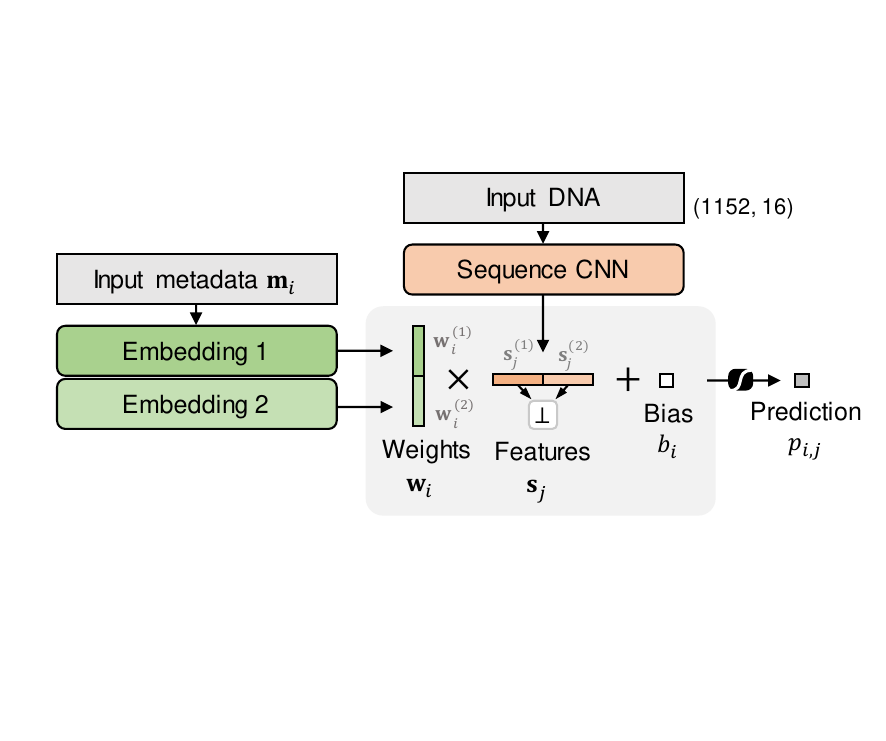}
  \caption{}
  \label{fig:model_schematic_1}
\end{subfigure}%
\begin{subfigure}{.5\textwidth}
  \centering
  \includegraphics[trim={0 6.5cm 0 0},clip,width=1\linewidth]{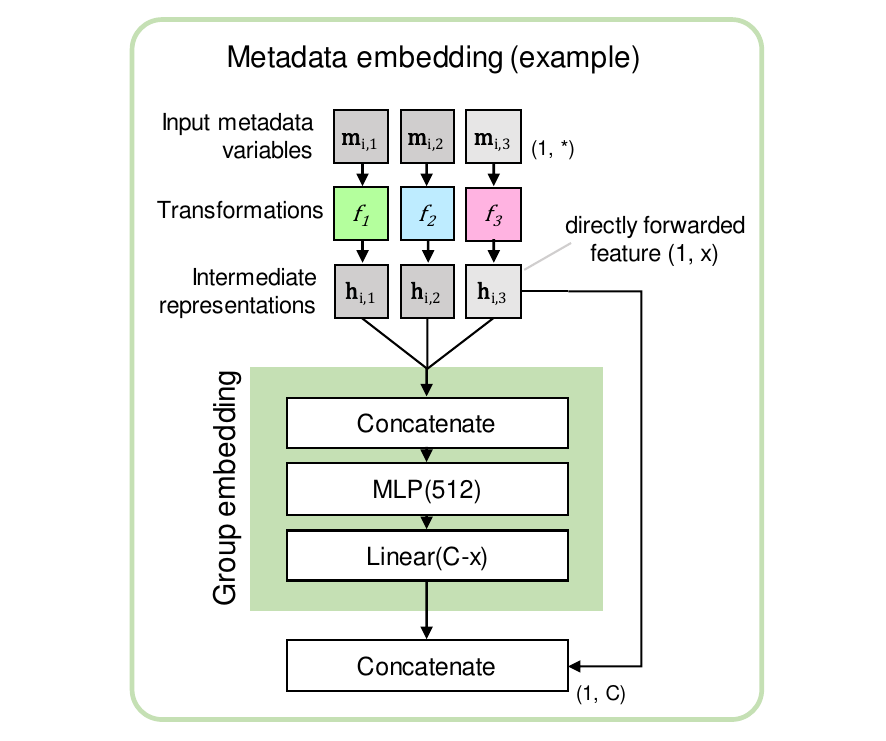}
  \caption{}
  \label{fig:model_schematic_2}
\end{subfigure}
    \caption[Model architecture and example of the metadata embedding module]{\textbf{Model architecture and example of the metadata embedding module.} \textbf{a)} Variables in metadata row $\mathbf{m}_i$ for class $i$ are fed through two metadata embedding modules 1 and 2 to produce weights $\mathbf{w}_i^{(1)}$ and $\mathbf{w}_i^{(2)}$, with $\mathbf{w}_i=[\mathbf{w}_i^{(1)}, \mathbf{w}_i^{(2)}]$. The sequence CNN extracts sequence features $\mathbf{s_j}$ from the 1152 bp sequence. Weights $\mathbf{w}_i$ are multiplied with $\mathbf{s_j}$, a bias $b_i$ is added, and the sigmoid activation function is applied to produce the prediction $p_{i,j}$. A penalty is placed on the \acl{MI} between features in $\mathbf{s}_j^{(1)}$ and $\mathbf{s}_j^{(2)}$ ($\perp$) in order to enforce independence between the two latent subspaces. \textbf{b)} A metadata embedding module with three variables $\textbf{m}_{i,1}$-$\textbf{m}_{i,3}$ (vectors or scalars), which are transformed by functions $f_{1}$-$f_{3}$ to produce intermediate variables $\mathbf{h}_{i,1}$-$\mathbf{h}_{i,3}$. The module can learn interactions between variables by feeding them through an MLP (Supplementary 
    Figure 5),
    followed by a linear mapping to $\latentSize-x$ dimensions. The metadata variable 3 with intermediate dimension $1 \times x$ is directly forwarded and concatenated to yield $\latentSize$ weights in total.}
\label{fig:funcgen_model_architecture_metadata_embedding_module}
\end{figure*}

We integrate the experiment metadata as follows: the metadata matrix $\mathbf{M} \in \R^{\numClasses \times \numMetadataVars}$ is non-linearly transformed via metadata embeddings -- trainable \acp{MLP} -- to derive weights of the output layer of the network $\mathbf{W} \in \R^{\latentSize \times \numClasses}$, where $\latentSize$ is the number of latent features from the sequence model, $\numMetadataVars$ is the number of metadata variables, and $\numClasses$ is the number of experiments.
To produce a single prediction $p_{i,j}$ for class $i$ and sequence $j$, the corresponding row in the weights matrix $\mathbf{w}_i$ is multiplied with the sequence representation $\mathbf{s}_j \in \R^{1 \times \latentSize}$, a class-specific bias ($b_i$) is added, and a sigmoid activation is applied (Figure \ref{fig:funcgen_model_architecture_metadata_embedding_module}):
\begin{equation}
\label{eq:model_per_class_prediction}
     p_{i,j} = \sigmoid \left(\mathbf{s}_j \mathbf{w}_i + b_{i} \right) 
\end{equation}
We divide the metadata variables into two groups, loosely interpretable as biological (e.g., tissue type, life stage, target) or technical (e.g., year, facility) experimental factors. The set of biological features is motivated by the fact that different tissues have distinct genetic programs that change during an organism's development and, therefore, will differ in epigenetic targets (e.g., whether DNA is accessible or if a repressive mark is present). In turn, technical features contain information about biases that arise from experimental procedures and batch effects. We train a separate embedding module $\MetadataEmbedding^{(i)} : \R^{\numMetadataVars^{(i)}} \mapsto \R^{\latentSize / 2}, i \in \{1, 2\}$ for each feature group. 
The two resulting sets of weights $\mathbf{w}^{(1)}$ and $\mathbf{w}^{(2)}$ are separately applied to the first and second halves of the extracted sequence features $\mathbf{s}_j^{(1)}$ and $\mathbf{s}_j^{(2)}$:
\begin{align}
\label{eq:model_per_class_prediction_disentangled}
      p_{i,j} =
      \sigmoid \left( \mathbf{s}_j^{(1)} \MetadataEmbedding^{(1)} \left(\mathbf{M} \right)_i  + \mathbf{s}_j^{(2)} \MetadataEmbedding^{(2)} \left(\mathbf{M} \right)_i  + b_{i} \right) = \nonumber \\
      \sigmoid \left( \mathbf{s}_j^{(1)} \mathbf{w}_i^{(1)}  + \mathbf{s}_j^{(2)} \mathbf{w}_i^{(2)}  + b_{i} \right)
\end{align}
This means that the biological metadata variables can only influence the final predictions via features from the first subset $\mathbf{s}^{(1)}$, while technical metadata can only utilize features from $\mathbf{s}^{(2)}$.
We further note that the metadata embeddings have an additional regularizing effect, as two classes with identical metadata are considered replicates, and share the same weights $w_i$ in the output layer - their predictions differ only by their class biases.
\subsection{Learning Independent Subspaces}
\label{ssec:model_disentanglement}
\newcommand{\Loss}{\mathcal{L}}
\newcommand{\LossAdv}{\Loss_{indep}}
\newcommand{\AdvPredictor}{\varrho}
\newcommand{\coeffCov}{\lambda_{indep}}
In order to learn disjoint features for the two latent subspaces, we additionally train the model to minimize the \ac{MI} between the biological and technical feature subspaces, using an adversarial training approach.
We train two \acp{MLP} models, denoted as $\AdvPredictor_{1-2}$ and $\AdvPredictor_{2-1} : \R ^{\latentSize/2} \mapsto \R^{\latentSize/2}$, to predict biological features from the technical ones, and vice-versa.
Specifically, during the adversarial training step we minimize:
\begin{align}
\label{eq:loss_fn_adv}
\LossAdv = - \sum_i^{\latentSize / 2} 
    \left[\rho_{(i)} \left( \mathbf{s}^{(2)}, \AdvPredictor_{1-2} (\mathbf{s}^{(1)}) \right)\right]^{2} - \nonumber \\
    - \sum_i^{\latentSize / 2} 
    \left[\rho_{(i)} \left( \mathbf{s}^{(1)}, \AdvPredictor_{2- 1} (\mathbf{s}^{(2)})
    \right)\right]^{2}
\end{align}
where with $\rho_{(i)}(\mathbf{x}, \mathbf{y})$ we denote the Pearson's correlation between the $i$-th dimensions of $\mathbf{x}$ and $\mathbf{y}$ computed empirically over a mini-batch of samples.
Consequently, the objective for the training step of the sequence model becomes:
\begin{align}
\label{eq:loss_fn}
\Loss_{\acs{MethodName}} = -\coeffCov \LossAdv~+ \nonumber\\
\frac{1}{N\numClasses}\sum_{j=1}^{\numClasses} \sum_{i=1}^{N} y_{i,j}\ \text{log}(p_{i,j}) + (1-y_{i,j})\ \text{log}(1-p_{i,j})
\end{align}
where $y_{i,j}$ is the binary label of the $j$-th class for the $i$-th mini-batch sample, and $\coeffCov$ controls the strength of the subspace-independence penalty.
Employing the independence penalty in the form of adversarially trained predictors, as opposed to, e.g., a cross-covariance penalty, ensures the independence of the subspaces in a general sense, constrained only by the capacity of $\AdvPredictor$, and not limited to simple linear dependencies (see Appendix 
Section B).
\section{Results}
Here we demonstrate how \ac{MethodName} allows for increased interpretability, by linking latent \ac{DL} features to different experimental factors (Section~\ref{ssec:interpretability}), while retaining or even improving performance on downstream tasks such as enhancer prediction (Section~\ref{ssec:enhancer}) and variant effect prediction (Section~\ref{ssec:vep}), as compared to a baseline model without metadata and independence constraints.
All the results are obtained with models pretrained on the \ac{ENCODE} data (Appendix 
Section A).
\subsection{\ac{MethodName} Enables Interpretation of Experimental Factors}
\label{ssec:interpretability}
To determine what the latent subspaces learned, we interpret the models by using Integrated Gradients \cite{sundararajan2017axiomatic}. To this end, we apply the neuron attribution implementation from the Captum package \cite{captum} to each node in the latent subspace layer to determine contribution scores for each position in the input sequence. Since the sequences are dinucleotide-encoded, we assign the contribution score to the first nucleotide of the two nucleotides, which corresponds to the nucleotide at the given position.

\begin{figure*}
\centering
\begin{subfigure}{.7\textwidth}
  \centering
  \includegraphics[width=1\linewidth]{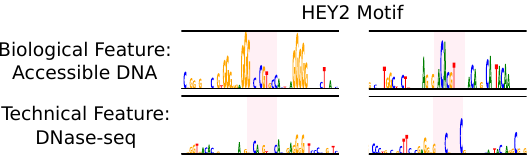}
  \caption{}
  \label{fig:hey2_motif}
\end{subfigure}%
\begin{subfigure}{.28\textwidth}
  \centering
  \includegraphics[width=1\linewidth]{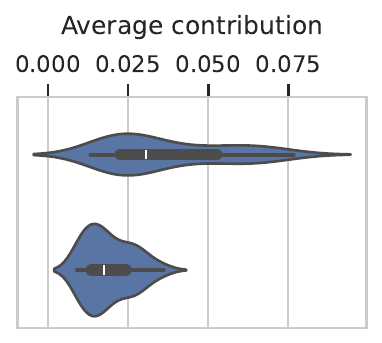}
  \caption{}
  \label{fig:avg_contrib}
\end{subfigure}
\caption{\textbf{Exemplary case of interpretability: contribution scores for direct features corresponding to 'Accessible DNA' (biological) and 'DNase-seq' (technical).} \textbf{(a)} The highlighted region represents the HEY2 motif for two sample sequences. \textbf{(b)} The average contribution scores for the context region ±50 bp around the motif for sequences with the HEY2 motif within the central 128 bp (n=40). We show how \ac{MethodName} allows for the interpretation and comparison of how input sequences interact with different experimental factors, using features directly corresponding to metadata factors. The `DNase-seq` feature is sensitive to different characteristics around the HEY2 motif than the `Accessible DNA` feature \textbf{(a)}, and is overall less influenced by the motif \textbf{(b)}.
}

\label{fig:contrib}
\end{figure*}

As an example case, we evaluate contribution scores for sequences with the HEY2 \ac{TF}-binding motif. HEY2 is known to be a regulator of early heart development. We select regions from test chromosomes that have HEY2 binding motifs and focus on the biological target feature, 'Accessible DNA,' and the technical feature, 'DNase-seq' (Figure \ref{fig:contrib}). DNase-seq is an experimental procedure to measure DNA accessibility or “openness” that is often interpreted as sequence activity. The motif is present in the attribution maps for the biological feature, while it cannot be observed in those for the technical feature for the same input sequence. The average contribution for sequences with the HEY2 motif within the central 128 bp window varies between the features. This indicates that the subspaces capture different signals and confirms that the 'Accessible DNA' feature attends to biologically meaningful motifs.

\begin{figure*}
\centering
\begin{subfigure}{.49\textwidth}
  \centering
  \includegraphics[width=1\linewidth]{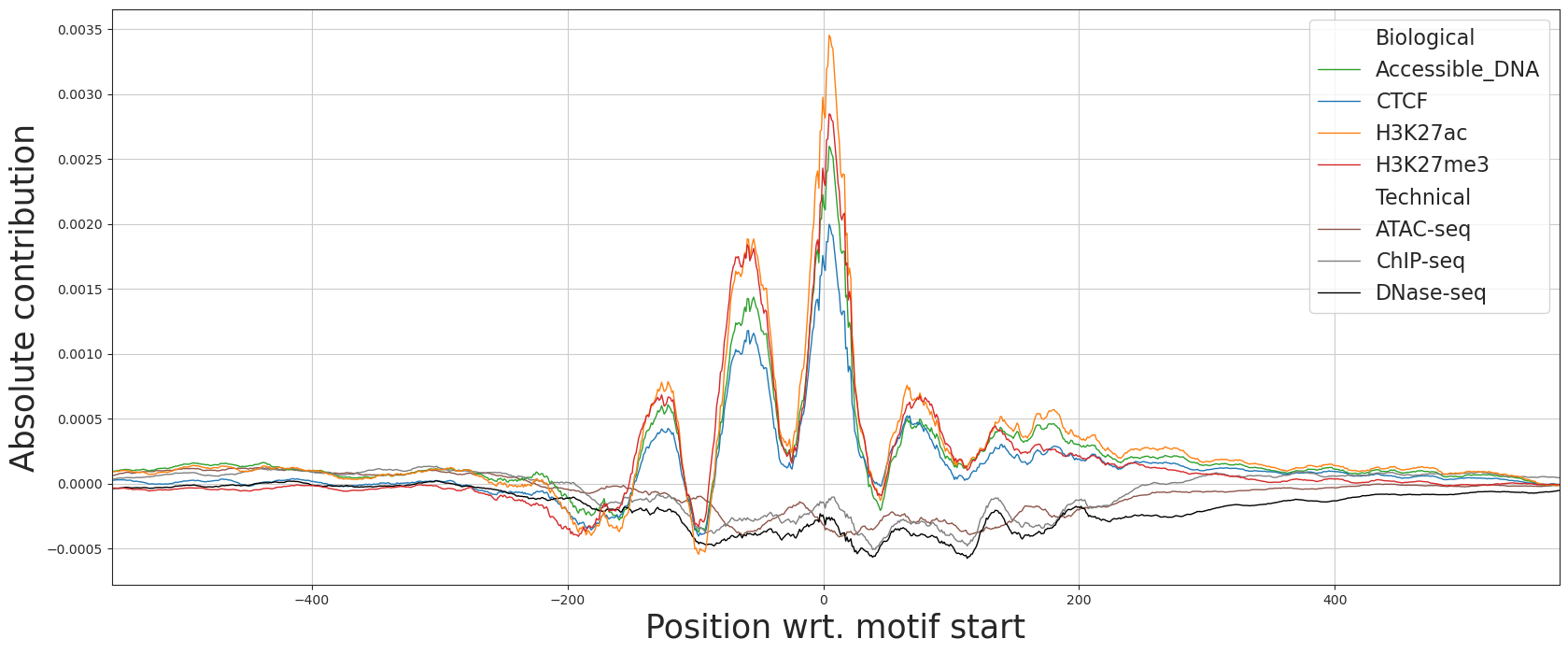}
  \caption{}
  \label{fig:heart_abs_sub}
\end{subfigure}%
\begin{subfigure}{.49\textwidth}
  \centering
  \includegraphics[width=1\linewidth]{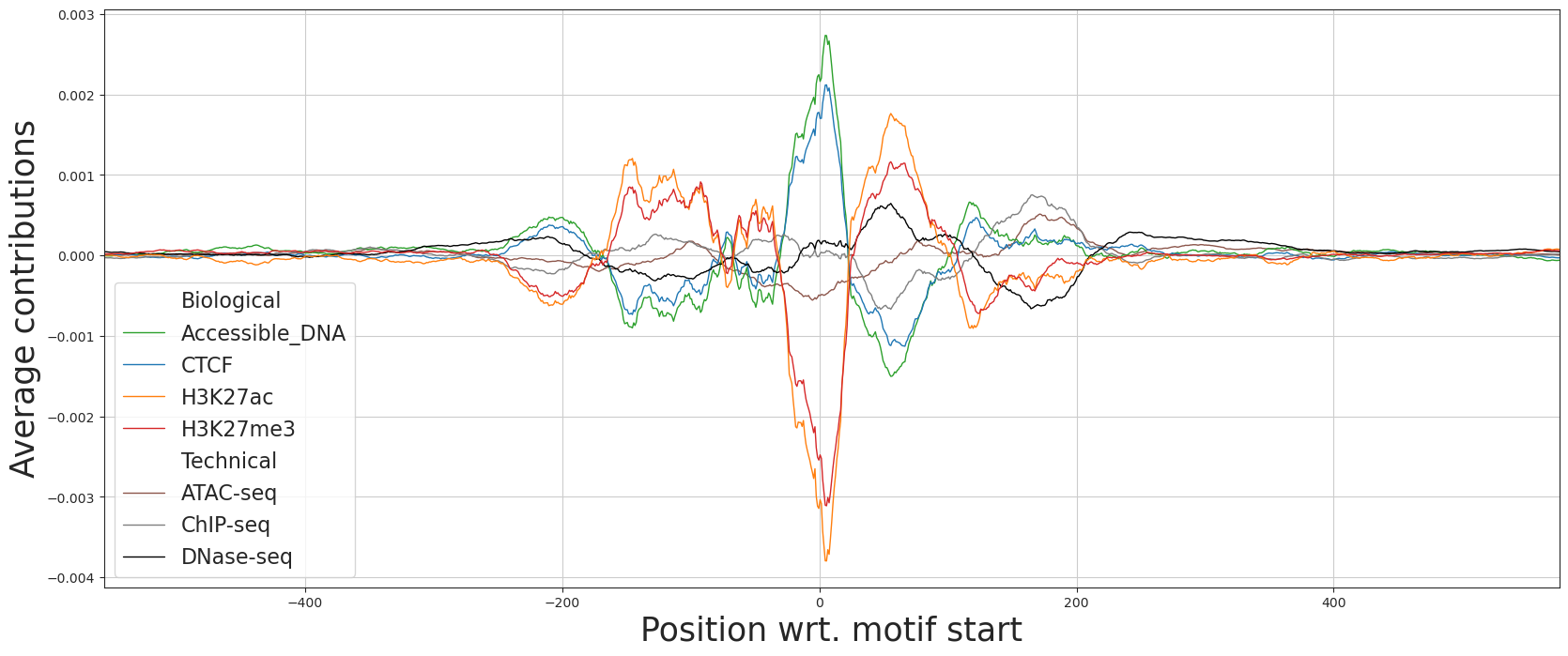}
  \caption{}
  \label{fig:heart_avg_sub}
\end{subfigure}
\begin{subfigure}{.49\textwidth}
  \centering
  \includegraphics[width=1\linewidth]{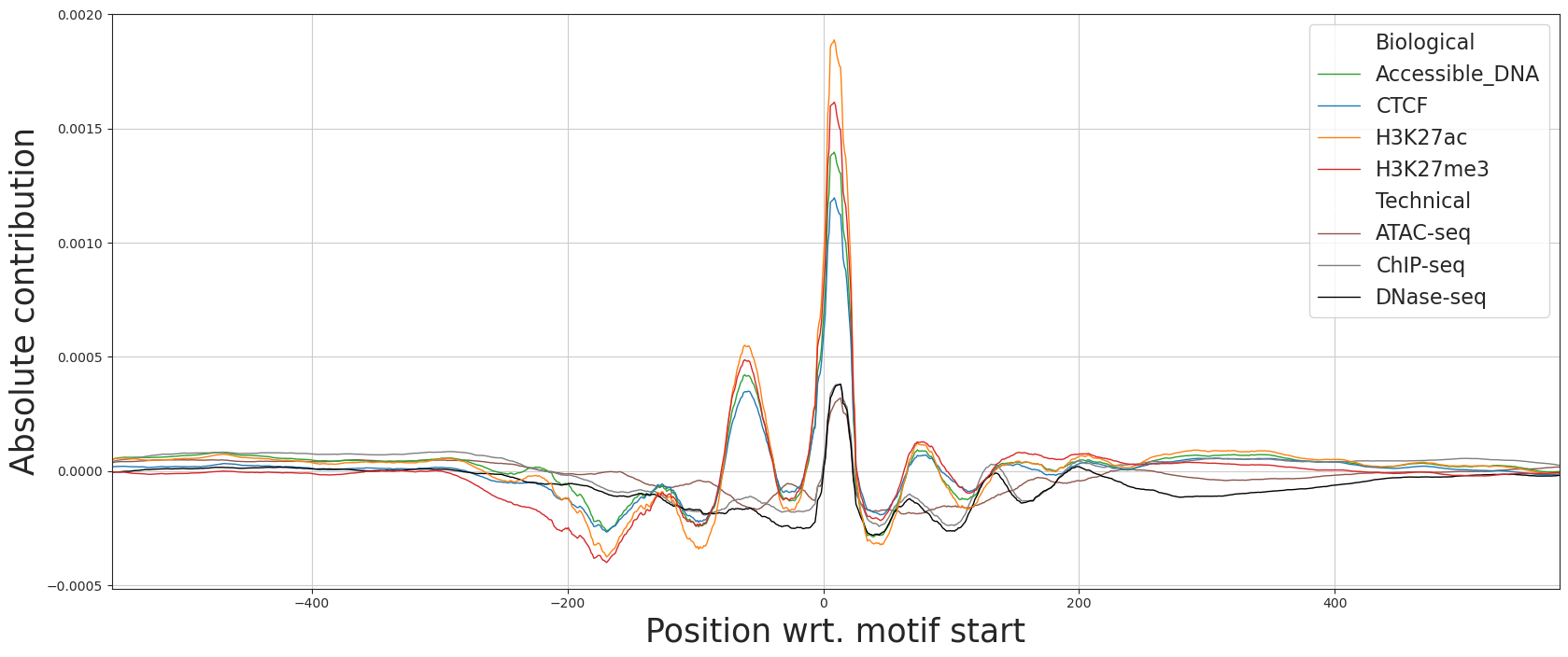}
  \caption{}
  \label{fig:ctcf_abs_sub}
\end{subfigure}%
\begin{subfigure}{.49\textwidth}
  \centering
  \includegraphics[width=1\linewidth]{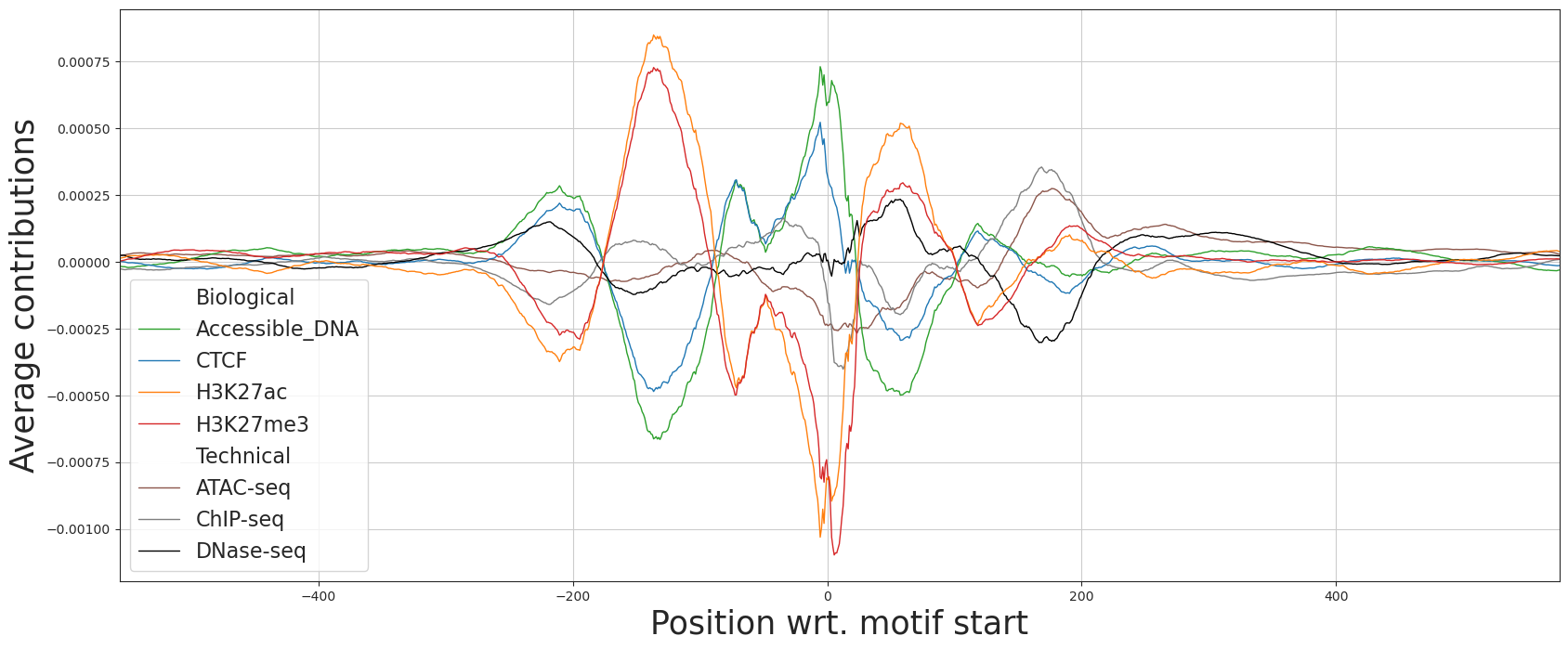}
  \caption{}
  \label{fig:ctcf_avg_sub}
\end{subfigure}
\caption{\textbf{Contribution scores wrt. position in input sequence for direct metadata features for targets and assays} \textbf{(a)} absolute and \textbf{(b)} average scores per base pair positions for 100 sequences with 
footprints for heart tissue
\textbf{(c)} absolute and \textbf{(d)} average scores per base pair positions for 4,457 sequences with CTCF motifs.
We subtracted a ``baseline'' signal from all examples, computed from $4,569$ sequences which had no corresponding experiment peaks.
Using features directly corresponding to metadata factors allows us to interpret model predictions on a finer scale.
For example, features corresponding to assay type seem to ignore the heart \acp{TF} motifs \textbf{(a)(b)}, while they seem sensitive to the CTCF ones \textbf{(c)(d)}, as indicated by the peaks around the start of the CTCF motifs.
Furthermore, the features of histone modifications (H3K27ac, H3K27me3) react in the opposite direction than features of CTCF and Acc. DNA \textbf{(b)(d)}.
}
\label{fig:contribs_wrt_pos_sub}
\end{figure*}

Furthermore, we examine attribution scores for sequences from test chromosomes with identified \ac{TF} footprints. Footprints were previously identified using DNase-seq experiments from ENCODE~\citep{Bentsen2020}. They indicate short 16 nucleotide-long regions of estimated \ac{TF} binding sites. We compute contribution scores for the directly forwarded features for targets (Accessible DNA, CTCF, H3K27ac, H3K27me3) and assays (DNase-seq, ATAC-seq, ChIP-seq) for two groups of sequences. The first group consists of 100 sequences, each centered on a unique high-score footprint with no other high-score footprints within 400 nucleotides upstream or downstream of the center (Figures~\ref{fig:heart_abs_sub},\ref{fig:heart_avg_sub}). The second group consists of $4,457$ sequences centered on footprints containing a CTCF binding motif (Figures~\ref{fig:ctcf_abs_sub},\ref{fig:ctcf_avg_sub}). 
CTCF is a ubiquitous transcription factor present in all cell types. 
We also calculate attribution scores for “baseline” sequences, defined as those exhibiting fewer than two signal peaks across all ENCODE experiments used as classes in training, and thus having no \ac{TF} footprints, and subtract them from the motif contributions, in order to separate motif-specific contributions from the baseline signal of the model 
(see Figure 7
of the Appendix for examples of the baseline signal and uncorrected plots).
Resulting plots show that the center of the sequence (the footprint) has high attribution scores for biological features such as Accessible DNA and CTCF and lower for technical features.
This suggests that latent biological features correspond to meaningful biological signals within the input sequences.
However, the observed periodical pattern, especially visible in the absolute contribution plots, might be an artifact of convolutional layers of the model.

\subsection{Biological Features Suffice for Enhancer Prediction}
\label{ssec:enhancer}
With trained DRL models at hand, we reason that the learned separation of latent subspaces into biological and technical can provide more robust features for downstream tasks. To evaluate this, we set up binary classification tasks to predict enhancer activity in the FANTOM5 dataset~\citep{dalby_2017_556775} and enhancer presence in the Vista dataset~\citep{visel2007vista}.
We encode the sequences using pretrained \ac{MethodName} models, obtaining 3 sets of features: biological $\featuresBio \in \R^{\numSamples \times \latentSize}$, technical $\featuresTech \in \R^{\numSamples \times \latentSize}$, and combined $\featuresFull = [\featuresBio, \featuresTech] \in \R^{\numSamples \times 2\latentSize}$. 
Each sequence is encoded in both the forward as well as the reverse directions, and the corresponding features are concatenated, resulting in $\latentSize$ features per subspace (instead of $\latentSize/2$).
Features obtained this way serve as inputs for regularized logistic regression models to predict the probability of a DNA sequence being an enhancer. 
For each tissue type, we train and evaluate 12 Ridge logistic regression models using \ac{MethodName} features: 3 feature types ($\featuresBio, \featuresTech, \featuresFull$) $\times$ 4 \ac{MethodName} models trained with different values of $\coeffCov$ 
(see Appendix E for more details).
Additionally, we evaluate features from a baseline model without metadata embeddings and independence constraints.

Within \ac{MethodName} features, the biological features achieve the highest mean \ac{AUROC} values in all but one setting (Table~\ref{tab:fantom} and 
Appendix Table 4).
We observe that both technical and biological features achieve comparable results, pointing to the worrisome scenario where predictions of classifiers that do not explicitly account for sources of noise may be based on artifacts rather than biology. However, our disentangled biological features do surpass the technical ones, and combining both feature subspaces does not yield better performance than the biological features alone.
This underlines the success of our DRL strategy and indicates that the biological features generalize better.
Furthermore, compared to a ``raw'', unregularized baseline model, \ac{MethodName} retains the predictive performance, while offering increased interpretability.

\begin{table}[t]
\caption{
Results of the enhancer classification task on the FANTOM5 dataset.
For each available tissue type we train a range of logistic regression models using different features obtained from pretrained \ac{MethodName} models and report mean \ac{AUROC} values computed across all tissue types.
We found that biological \ac{MethodName} features alone are as predictive as features from an unconstrained baseline model.
}
\label{tab:fantom}
\begin{center}
\begin{tabular}{l|ccc}
\hline
 $\coeffCov$ &  Combined & Biological & Technical  \\
\hline
       Baseline &  \textbf{0.68} &       - &      -  \\
        0 &  0.66 &       0.67 &      0.66\\
       0.001 &  0.67 &       \textbf{0.68} &      0.62 \\
       0.01 &  0.63 &       0.64 &      0.61  \\
       0.1 &  0.58 &       0.58 &      0.56  \\
\hline
\end{tabular}
\end{center}
\end{table}
\subsection{Biological Features Improve Variant Effect Prediction}
\label{ssec:vep}
We further evaluate the utility of \ac{MethodName} in a zero-shot variant effect prediction task.
Selecting the model pretrained with $\coeffCov$ of $0.001$, based on its performance in the enhancer prediction task on the FANTOM5 dataset (Section~\ref{ssec:enhancer}),
we encode for each variant its corresponding reference and alternative sequences, obtaining features $\featuresFull^{ref} = [\featuresBio^{ref}, \featuresTech^{ref}]$ and $\featuresFull^{alt} = [\featuresBio^{alt}, \featuresTech^{alt}]$.
\acp{VEP} are then calculated as the difference in model predictions: $\predFn (\featuresFull^{ref}) - \predFn (\featuresFull^{alt})$, where $\predFn_i(\mathbf{x}) = \sigmoid(\mathbf{x}\mathbf{w}_i + b_i)$ 
is the prediction for the \textit{i}-th output class
(see Equation~\ref{eq:model_per_class_prediction} and Figure~\ref{fig:funcgen_model_architecture_metadata_embedding_module}). 
We further obtain \acp{VEP} for the biological signal by calculating predictions for the alternative allele as $\predFn([\featuresBio^{alt}, \featuresTech^{ref}])$, i.e., using biological features for the alternative allele sequence and technical features for the reference one (and vice versa for the technical \acp{VEP}).
We average model predictions across small shifts around the center and average the predictions for the forward and reverse strands.

\begin{figure*}
\centering
\begin{subfigure}{.5\textwidth}
  \centering
  \includegraphics[width=1\linewidth]{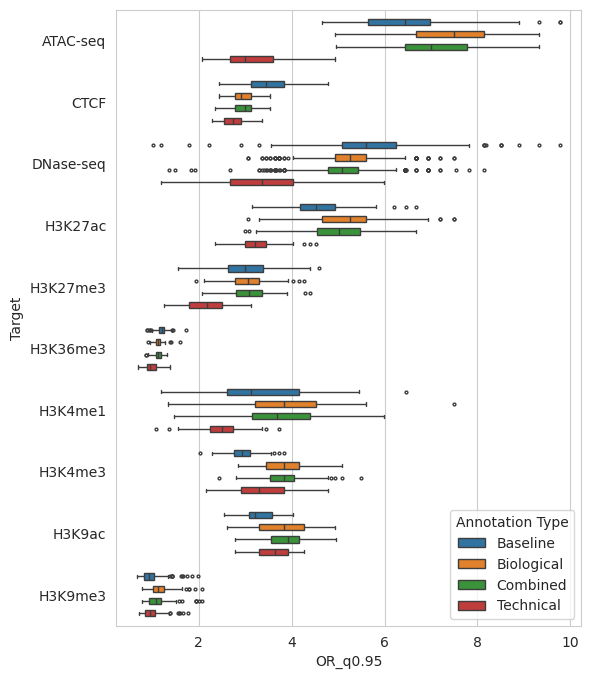}
  \caption{}
  \label{fig:gtex_or_09}
\end{subfigure}%
\begin{subfigure}{.5\textwidth}
  \centering
  \includegraphics[width=1\linewidth]{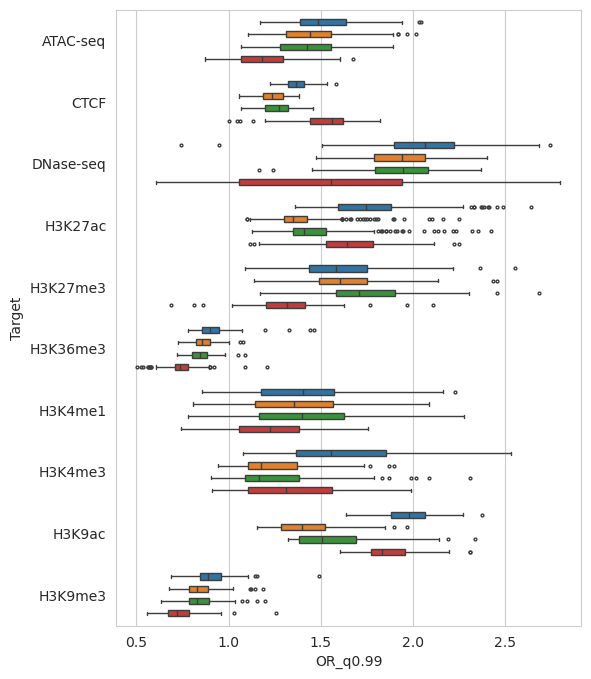}
  \caption{}
  \label{fig:gnomad_pls_or_09}
\end{subfigure}
\caption{
\textbf{Mean odds ratios of identifying:} \textbf{a)} an \ac{eQTL} variant in the \ac{GTEx} dataset
and \textbf{b)} a rare variant in the gnomAD dataset, over different assays and targets for different feature types.
We use the upper 95-th and 99-th quantiles as the cutoff for identifying the variants for \ac{GTEx} and gnomAD respectively.
}
\label{fig:vep_or_09}
\end{figure*}

By choosing a cutoff value based on the quantiles of the resulting distribution of \acp{VEP}, we perform zero-shot variant discoveries for \ac{eQTL} variants in the \ac{GTEx}~\citep{lonsdale2013genotype}, and rare PLS-CRE variants in the gnomAD~\citep{benegas2023gpn} datasets, which we describe in more detail in Appendix 
Sections F.1 and F.2.
We compute the overall enrichment per \ac{VEP} annotation type by aggregating the tagged variants across all $\numExperimentsNumeric$ outputs (Table~\ref{tab:vep}).
For the first two quantile cutoffs, all feature types yield comparable \acp{OR}; for the most extreme cutoffs, the technical annotations achieve a $7\%$ lower enrichment for both datasets.
Features from the baseline model yield no improvement over the combined or biological ones in all the settings.
Overall, the biological annotations yield an improvement over the baseline in all quantile settings in both datasets.
To gain insights into potential class biases, we plot the mean \acp{OR} using \acp{VEP} corresponding to predictions within each target assay 
in Figure~\ref{fig:vep_or_09}.
The combined and biological \acp{VEP} consistently yield comparable enrichment values, while the technical ones vary more strongly across targets.
\begin{table}[t]
\caption{
\textbf{Enrichments of:} \textbf{a)} an \ac{eQTL} variant in the \ac{GTEx} dataset and \textbf{b)} a rare variant in the gnomAD dataset, over all experiment outputs per-feature (combined, biological, and technical predictions). 
The values are computed over the total numbers of unique true positive and false positive variants identified.
\ac{MethodName}-derived features improve performance over the baseline, while allowing for greater interpretability - separating the biological and technical factors shows that albeit the technical features are predictive, the biological ones alone suffice for good performance.
}
\label{tab:vep}
    \begin{subtable}{1\linewidth}
      \centering
        \caption{}
\begin{tabular}{llr}
\hline
Quantile & Annotation &     Enrichment              \\
\hline
0.9 & Baseline &              1.04 \\
      & Biological &              1.05 \\
      & Combined &              1.05 \\
      & Technical &              1.05 \\
0.95 & Baseline &              1.12 \\
      & Biological &              1.13 \\
      & Combined &              1.12 \\
      & Technical &              1.14 \\
0.99 & Baseline &              1.42 \\
      & Biological &              1.43 \\
      & Combined &              1.42 \\
      & Technical &              1.32 \\
\end{tabular}
    \end{subtable}%
    \\
    \begin{subtable}{1\linewidth}
      \centering
        \caption{}
\begin{tabular}{llr}
\hline
Quantile & Annotation &   Enrichment                \\
\hline
(0.1, 0.9) & Baseline &              1.15 \\
                     & Biological &              1.16 \\
                     & Combined &              1.16 \\
                     & Technical &              1.16 \\
(0.01, 0.99) & Baseline &              1.27 \\
                     & Biological &              1.28 \\
                     & Combined &              1.29 \\
                     & Technical &              1.26 \\
(0.001, 0.999) & Baseline &              1.71 \\
                     & Biological &              1.77 \\
                     & Combined &              1.78 \\
                     & Technical &              1.66 \\
\end{tabular}
    \end{subtable} 
\end{table}
\section{Discussion}
\ac{MethodName} is a deep learning model designed to learn a disentangled representation of the human epigenome, trained to isolate low-dimensional biological features from those of a technical nature. On several independent downstream tasks, we demonstrated that predictive models utilizing the biological features outperform those that incorporate technical features or a combination thereof. This finding substantiates the model's capability to effectively separate technical biases inherent in the training data from genuine biological signals, thereby enhancing the accuracy of DNA sequence-based predictions through effective "de-noising."
The task of enhancer prediction presented a considerable challenge, primarily due to the complex and nuanced nature of gene regulation syntax. This complexity is reflected in the sub-optimal average AUROCs observed for enhancer classification tasks. Nevertheless, we demonstrated that \ac{MethodName}-derived biological features are sufficient to achieve the predictive performance of an unconstrained baseline model while offering greater interpretability.
In the variant effect prediction task, features derived from diverse experiments demonstrated variable success in identifying true variants, underscoring the profound impact of technical biases on prediction outcomes. However, when quantifying the overall enrichment, the \ac{MethodName} biological features consistently yielded better performance than the baseline model. Despite the considerable predictive power of technical features in several cases, we argue in favor of utilizing disentangled biological representations. By investigating model attribution maps, we showed how biological features attend to meaningful information (e.g., transcription factor motifs) in a DNA sequence, in contrast to the unspecific attributions for technical features.

\section{Funding}
This work was supported by the European Commission [101016775]; the Deutsche Forschungsgemeinschaft (DFG) [LI 3333/5-1, GR 3793/6-1, RE3474/8-1,OH 266/6-1]; the HPI Research School on Data Science and Engineering; and the Helmholtz Einstein International Berlin Research School in Data Science (HEIBRiDS) program of the Helmholtz Association.
We thank Bernhard Renard for insightful comments on the manuscript.
\bibliographystyle{bibstyle}
\bibliography{reference}
\clearpage
\begin{appendices}
\section{Model Training}
\label{app:training}
\begin{figure*}[!h]
    \centering
    \includegraphics[width=0.65\textwidth]{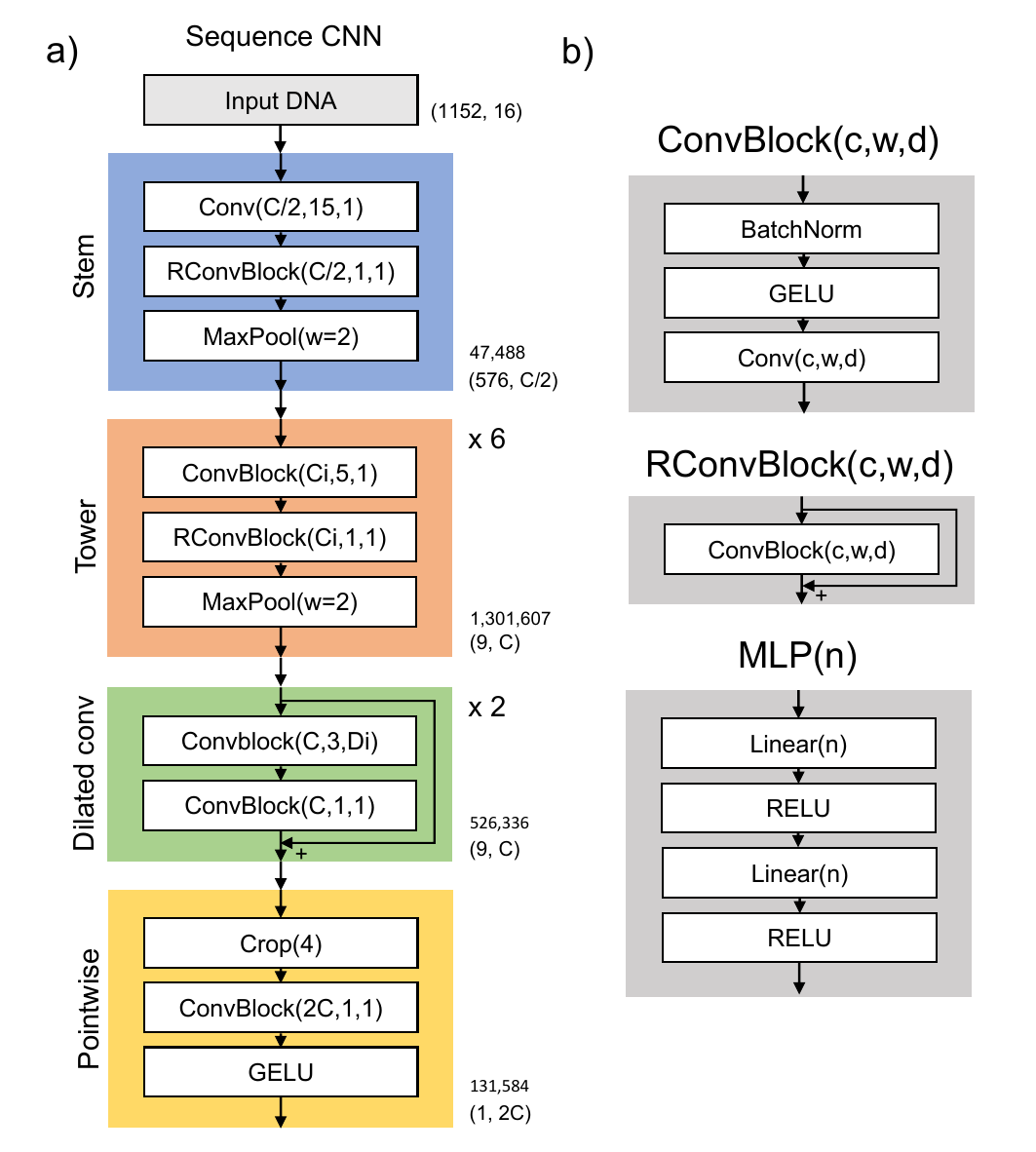} 
    \caption[DNA-sequence convolutional neural network architecture and building blocks]{\textbf{DNA-sequence convolutional neural network architecture and building blocks.} This figure has been adapted from ref \cite{avsec2021effective}, where it was published under CC-BY 4.0 license (\url{https://creativecommons.org/licenses/by/4.0/}). Changes were made to reflect the parameters used and the new MLP building block. \textbf{a)} The DNA-sequence CNN is divided into four modules. The stem acts as a short motif-finder ($\mathrm{C}/2$ channels). The tower grows the number of channels to $\mathrm{C}$ and reduces the spatial dimension/resolution. The dilated convolutions aggregate context across the sequence. The pointwise convolution transforms the sequence to its final intermediate representation with $\mathrm{2C}$ channels. The numbers in brackets next to boxes denote the dimensions (the first is the spatial dimension, the second the number of channels). The number of parameters of each module is shown above the brackets. Panel \textbf{b)} shows the implementation of the building blocks of panel a). BatchNorm: Batch normalization, GELU: Gaussian Error Linear Unit, RELU: Rectified Linear Unit, MLP: Multilayer Perceptron, MaxPool(w): MaxPooling with stride and width w, Conv(c,w,d): convolution with c channels, kernel width w, and dilation d. Linear(n): Fully connected layer with n outputs. All experiments used $\mathrm{C} = 128$.}
    \label{fig:sufpig_seq_cnn_architecture_and_building_blocks}
\end{figure*}

We train \ac{MethodName} models on dinucleotide sequences of length $1152$ from the GRCh38 human reference genome data, to predict peak-calls in $\numExperimentsNumeric$ tissue-specific DNA-accessibility (ATAC-seq, DNase-seq) and chromatin modification (ChIP-seq) experiments on human samples from the \ac{ENCODE} database.
We left out data from the 9th and 10th chromosomes as test data, and take $5\%$ of the remaining samples as validation data.
We augment the training data by randomly sampling either the forward or reverse complement of sequences, and applying random shifts of up to 8bp in either direction~\citep{kelley2018sequential,avsec2021effective}.
The models are optimized for 100 epochs using the AMSGrad variant of the Adam optimizer~\citep{reddi2019convergence,kingma2014adam} with a mini-batch size of 4096 and a learning rate of $10^{-3}$.
We monitor the \ac{AUROC} values of validation set predictions after each training epoch, and use the model weights with highest \ac{AUROC} values for downstream tasks.
We set $\latentSize=128$ as the dimensionality of the latent subspaces of the biological and technical features, and train 3 model variants, with the regularization coefficients for the subspace independence penalty $\coeffCov \in \{10^{-3}, 10^{-2}, 10^{-1} \}$.
The metadata variables used for training the models are listed in Table~\ref{tab:metadata_config}.
The backbone model for sequence feature extraction was based on the Basenji2 architecture~\citep{avsec2021effective}
(Figure~\ref{fig:sufpig_seq_cnn_architecture_and_building_blocks}).
Model definition and training were implemented using the PyTorch and Pytorch-Lightning frameworks~\citep{paszke2019pytorch,Falcon_PyTorch_Lightning_2019}.

\begin{table*}[t]
\caption{
Metadata features from \ac{ENCODE} experiments used for training of \acf{MethodName} models.
}
\label{tab:metadata_config}
\begin{center}
\begin{tabular}{l|l|l|l}
Feature & Direct & Interact. & Example Values  \\
\hline
Biosample Term & No  & Yes  & chorionic villus, right lung   \\
Biosample Organ & No  & Yes & intestine, spleen \\
Biosample Life Stage & No  & Yes &  adult, embryonic \\
Age & No  & Yes &  10, 55 \\
Age Unit & No  & Yes &  week, year \\
Target & Yes & Yes  & Acc. DNA, H3K27ac \\
Assay & No  & Yes &  Dnase-seq, ATAQ-seq \\
GC-mean & Yes  & Yes &  $0.43$, $0.57$ \\
Lab & No  & Yes & Bing-Ren, Bernstein \\
Year Released & No  & Yes & 2013, 2016 \\
\end{tabular}
\end{center}
\end{table*}
\section{Evaluating Subspace Independence}
\label{app:independence}
\begin{figure}[!h]
    \centering
    \includegraphics[width=0.5\textwidth]{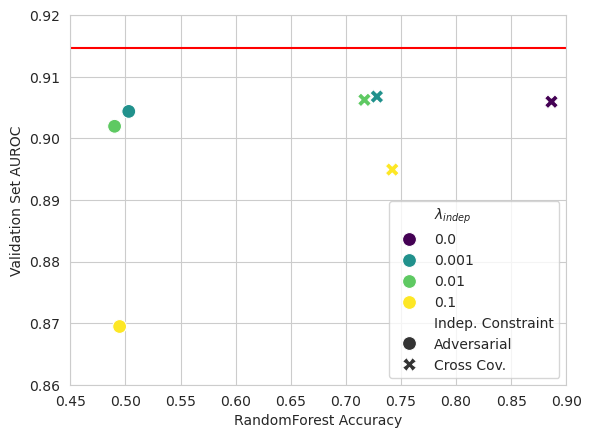}
    \caption{
    Comparison of independence between latent subspaces (x-axis), measured by the accuracy of predicting between pairs of original and shuffled batches, and model predictive performance (y-axis) as AUROC of validation samples, for different regularization strengths $\coeffCov$ and independence constraints.
    The horizontal red line indicates the performance of a baseline model, without metadata embeddings and independence constraints.
    }
    \label{fig:independence}
\end{figure}

We compare the effect of enforcing independence between the subspaces with an adversarial predictor (Section~\ref{ssec:model_disentanglement}), which can capture non-linear dependencies, to a linear constraint, which penalizes the Frobenius norm of the cross-covariance matrix between the two features sets:
\begin{align}
\mathcal{L}_{lin. indep.} = \lVert\text{cov}(\mathbf{s}^{(1)}, \mathbf{s}^{(2)}) \rVert_F
\end{align}
To quantify independence, we employ a batch-shuffling approach~\cite{belghazi2018mine} by training a Random Forest classifier to  distinguish between pairs of $(\mathbf{s}_i^{(1)}, \mathbf{s}_i^{(2)})$ and $(\mathbf{s}_i^{(1)}, \mathbf{s}_{shuff(i)}^{(2)})$, where $\mathbf{s}_{shuff(i)}^{(2)}$ contains elements of $\mathbf{s}^{(2)}$ with a randomly shuffled order of rows (observations).
By shuffling the order, we simulate drawing samples from $\mathbf{s}^{(2)}$ independently of $\mathbf{s}^{(1)}$.
If $\mathbf{s}^{(1)} \perp \mathbf{s}^{(2)}$, then we would have $P(\mathbf{s}_i^{(1)}, \mathbf{s}^{(2)}) = P(\mathbf{s}^{(1)})P(\mathbf{s}^{(2)})$.
We compare the achieved Random Forest accuracies and the AUROC scores of the corresponding model predictions for a set of models trained with different $\coeffCov$ values and independence constraints (Figure~\ref{fig:independence}).
All models trained with the adversarial constraint achieved a RandomForest accuracy of 50\%, meaning the classifier was not able to distinguish between the original and shuffled subspaces.
Enforcing just a linear independence resulted in an increase of the score to 75\%, whereas a model without any constraint ($\coeffCov = 0$) had a score of almost 90\%.
For $\coeffCov < 0.1$ the adversarial penalty achieved comparable AUROC performance to both the linear and unconstrained model.
Instead, we found a larger drop in AUROC, as compared to the baseline model,  stemming from employing the metadata embeddings themselves.

\section{Querying and processing ENCODE data}
\label{app:encode_get_data}

We queried the ENCODE database on the 27th of May 2021 and identified peak-calls from tissue-based experiments in human (assmembly GRCh38) samples. We considered only DNase-seq, ATAC-seq, CTCF ChIP-seq, and ChIP-seq for histone modifications (H3K27ac, H3K27me3, H3K9me3, H3K36me3, H3K4me1, H3K4me2, H3K4me3). We selected (pseudo-)replicated peaks for ATAC seq, IDR-thresholded peaks for CTCF and (pseudo-)replicated peaks for histone modifications. For DNase-seq, we considered all available peak files for a given experiment accession (with reported FDR = 0.05) because ENCODE did not provide (pseudo-)replicated or IDR-thresholded peaks for these experiments. Peak files were downloaded in narrowPeak format. 

We queried metadata (e.g., sample quality metrics or ontologies) for all experiments using the ENCODE REST API. For experiment we queried attributes of the linked Biosample, Library and Experiment objects. A Biosample relates to a unique sample of biological material. A Library is a unique sequencing library (a sample of processed DNA for sequencing), and an Experiment encompasses a group of one or more experimental replicates. From the Biosamples, we queried the life\_stage, age, and age\_units attributes. We further retrieved standardized ontological terms describing the tissue for each experiment, specifically the term\_name (e.g., "heart right ventricle") and the organ\_slims (e.g., "heart"). These attributes are hierarchical, i.e., multiple term\_names may map to the same organ\_slim, and a term name may have more than one organ\_slim (e.g., "intestine,large intestine"). Additionally, we queried metadata related to sample quality (e.g., the reported fraction of reads in peaks (FRIP)), the lab that produced the data, and the date the experiment was released.

We defined genomic regions of interest based on a set of peak files. This strategy was designed to be robust to within-sample outliers (extremely broad peaks) and between-sample outliers (extreme number of peaks). Processing is performed within groups (DNase, ATAC, separate histone modifications, CTCF). First, only peaks on autosomes and chromosome X are retained and peaks overlapping blacklisted regions are excluded (these include the ENCODE blacklist~\citep{amemiyaENCODEBlacklistIdentification2019} and a small set of Vista enhancers~\citep{viselVISTAEnhancerBrowser2007} used in downstream tasks). For each peak file, we calculated $w_{\text{max}}$ as the 75th percentile plus 1.5 times the IQR of the peak width. If that value was shorter than 1000, it was set to 1000. Peaks that were longer than $w_{\text{max}}$ were clipped so that their start and end coordinates did not extend any further than $w_{\text{max}}$ away from the reported peak center (this caps the maximum peak length at $2 w_{\text{max}}$). For each peak file, we calculated $w_{\text{max}}$ as the 75th percentile plus 1.5 times the IQR of peak width, with a minimum value of 1000. Peaks exceeding $w_{\text{max}}$ were trimmed to a maximum length of $2 w_{\text{max}}$ from the peak center. Similarly, for each sample group, $p_{\text{max}}$ was determined as the 75th percentile plus 1.5 times the IQR of peak counts. Files exceeding $p_{\text{max}}$ peaks were reduced to only the strongest $p_{\text{max}}$ peaks based on signal value.

For every sliding window within a set of regions of interest, I calculated overlaps to the original (i.e., non-filtered) peaks. Each peak file is considered its own class. If a region of interest is covered at least 50\% by a peak from a specific file (i.e., experiment/replicate), it is considered a positive for that class (1), otherwise it is considered a negative (0)~\citep{zhou2015deepsea}.


\section{Interpretability}
\label{app:interpretability}
\begin{figure*}
\centering
\begin{subfigure}{.49\textwidth}
  \centering
  \includegraphics[width=1\linewidth]{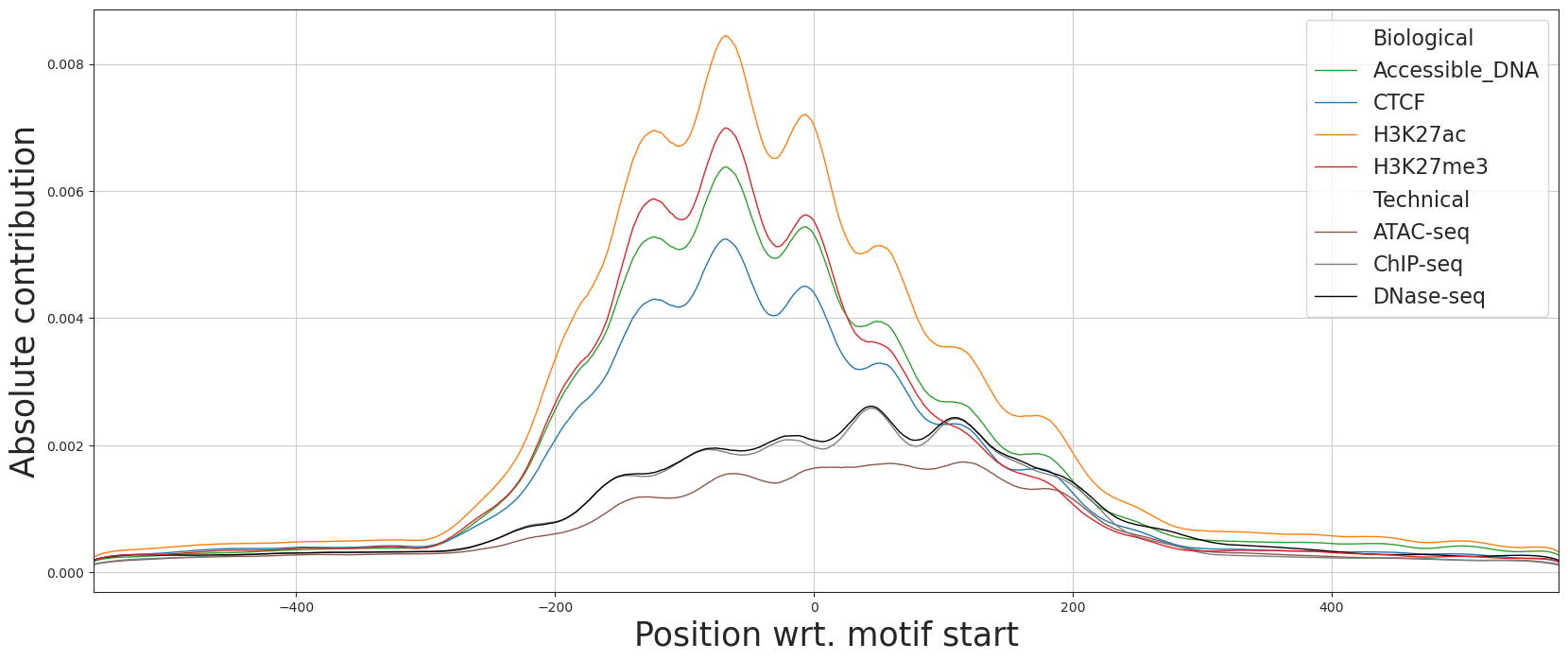}
  \caption{}
  \label{fig:neg_abs}
\end{subfigure}%
\begin{subfigure}{.49\textwidth}
  \centering
  \includegraphics[width=1\linewidth]{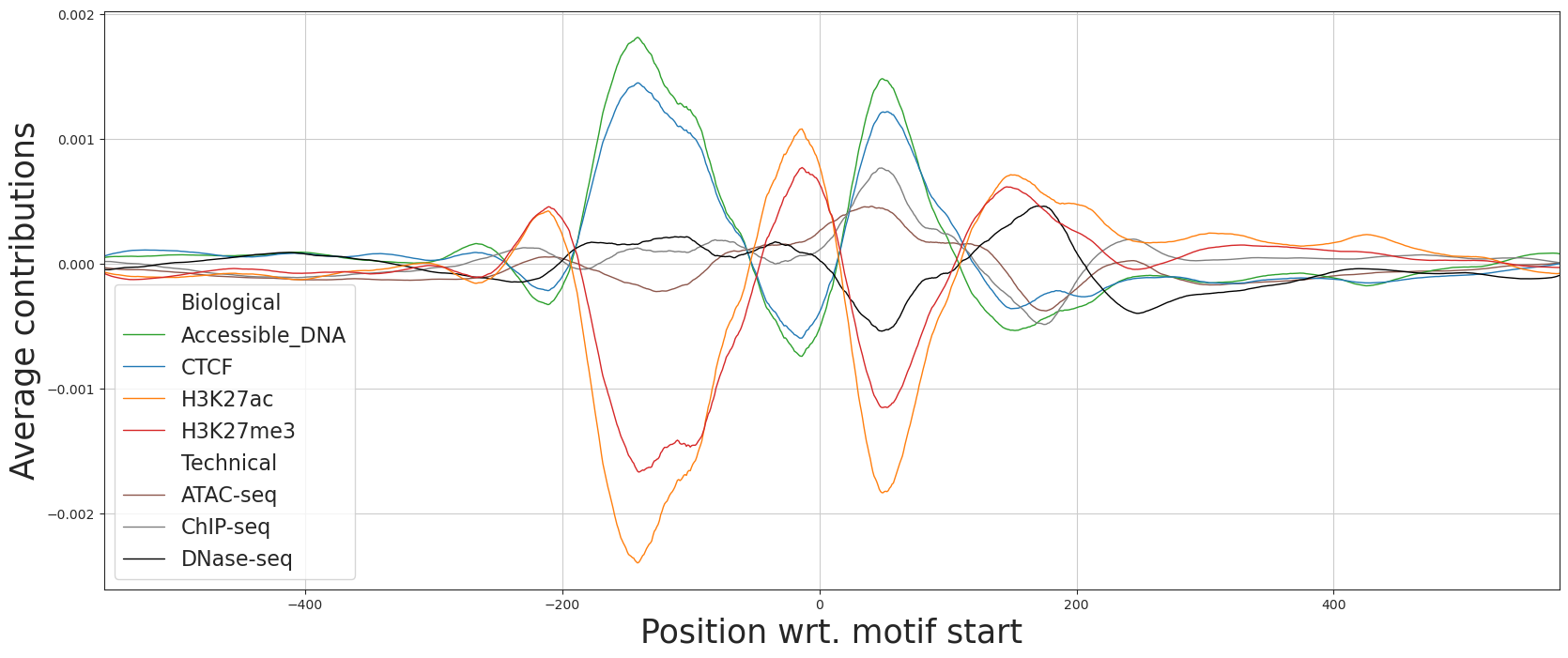}
  \caption{}
  \label{fig:neg_avg}
\end{subfigure}
\begin{subfigure}{.49\textwidth}
  \centering
  \includegraphics[width=1\linewidth]{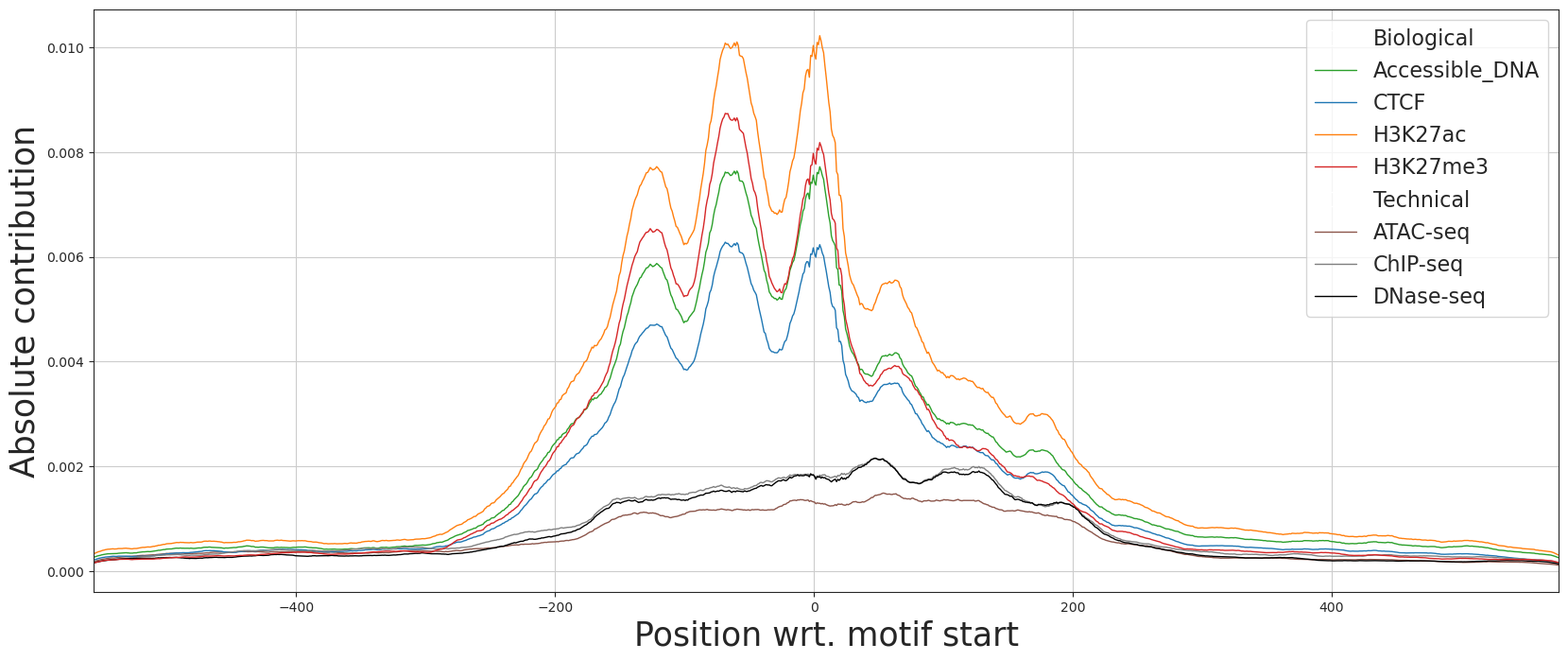}
  \caption{}
  \label{fig:heart_abs}
\end{subfigure}%
\begin{subfigure}{.49\textwidth}
  \centering
  \includegraphics[width=1\linewidth]{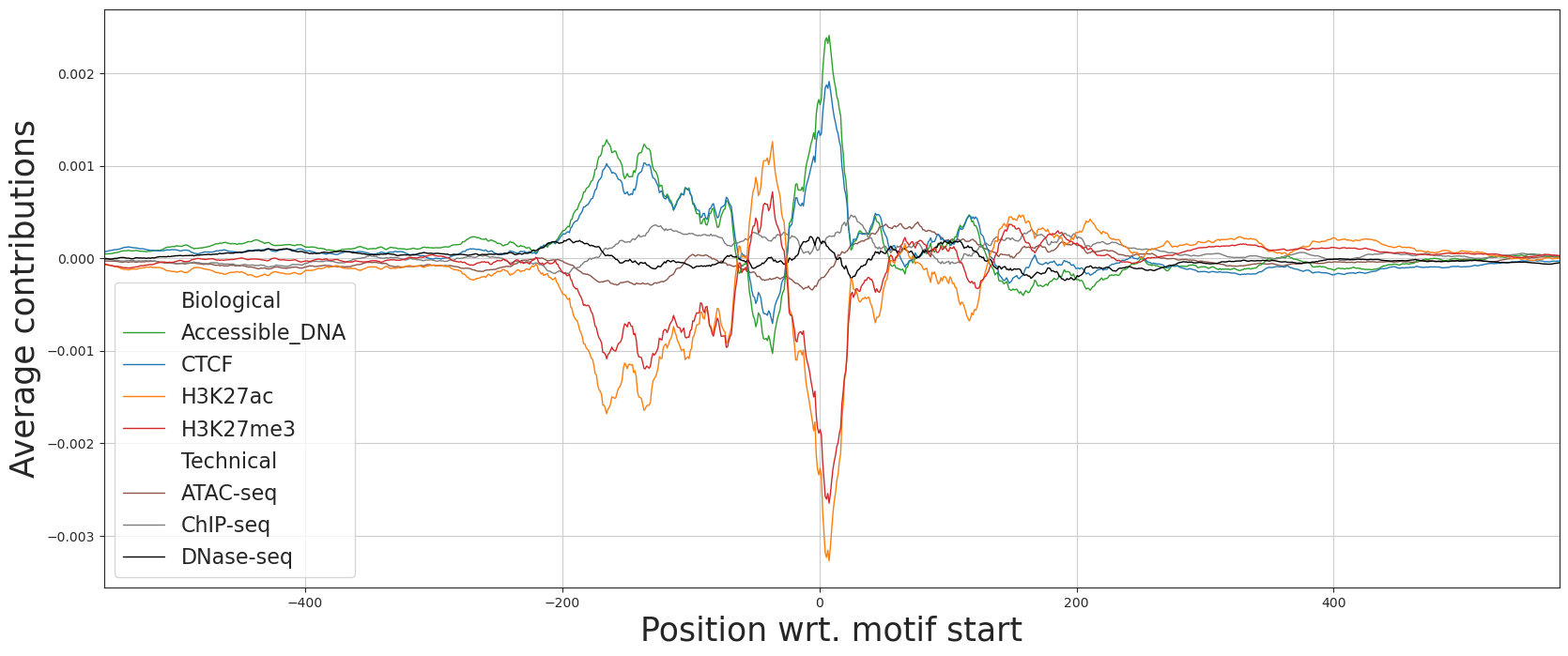}
  \caption{}
  \label{fig:heart_avg}
\end{subfigure}
\begin{subfigure}{.49\textwidth}
  \centering
  \includegraphics[width=1\linewidth]{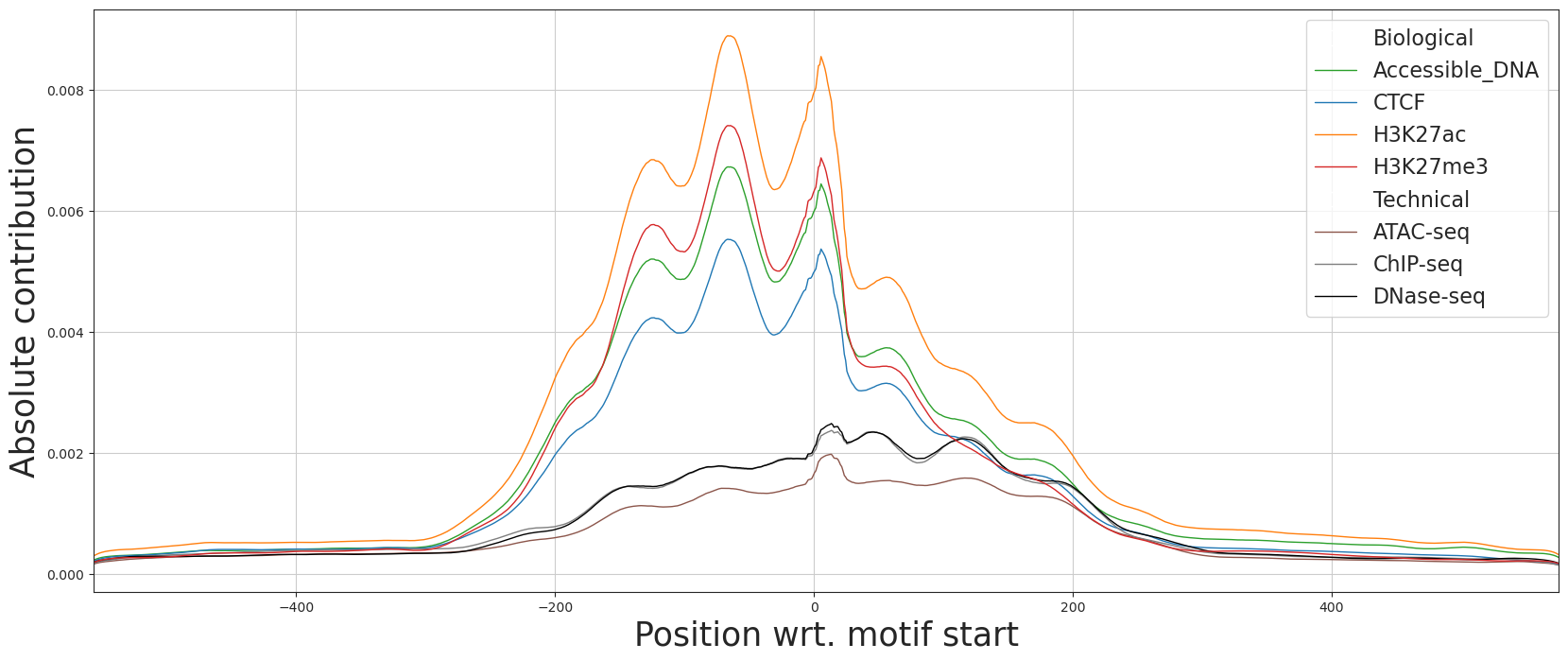}
  \caption{}
  \label{fig:ctcf_abs}
\end{subfigure}%
\begin{subfigure}{.49\textwidth}
  \centering
  \includegraphics[width=1\linewidth]{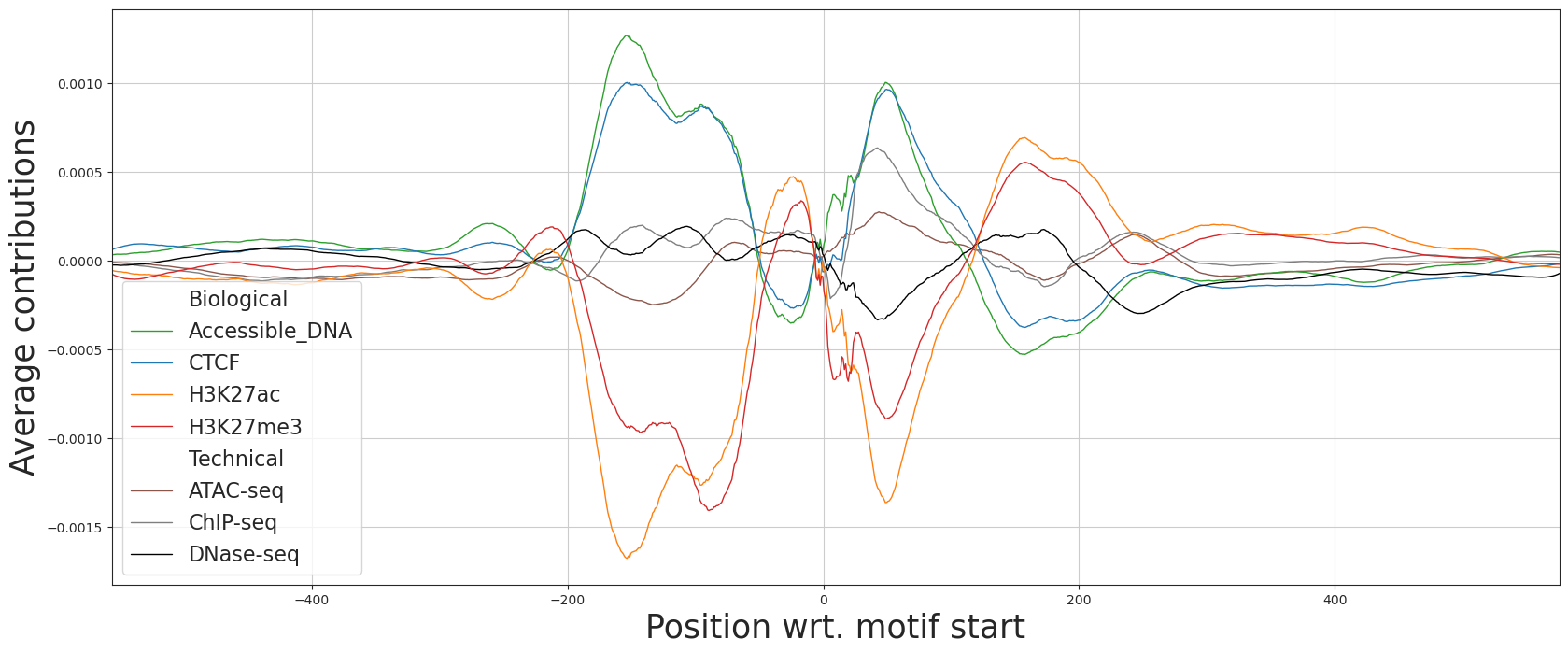}
  \caption{}
  \label{fig:ctcf_avg}
\end{subfigure}
\caption{
\textbf{Contribution scores for direct features for targets and assays} 
\textbf{(a)} absolute and \textbf{(b)} average scores per base pair positions for 4,569 sequences of negative samples
\textbf{(c)} absolute and \textbf{(d)} average scores per base pair positions for 100 sequences with \ac{TF}
footprints for heart tissue
\textbf{(e)} absolute and \textbf{(f)} average scores per base pair positions for 4,457 sequences with CTCF motifs.
In all cases we observe periodicity of peaks and asymmetry wrt. the center of the sequence, which we attribute to the workings of the underlying \ac{CNN} model due to their prevalence across all input types.
}
\label{fig:contribs_wrt_pos}
\end{figure*}

Figure~\ref{fig:contribs_wrt_pos} shows absolute (left column) and average (right column) contributions wrt. the center of the input sequence per target and assay type, for random negative sequences from the training data (first row), sequences centered around heart \ac{TF} footprints (second row), and CTCF motifs (third row).
While the features react differently to the different input sequences, they all exhibit asymmetrical behavior wrt. the sequence center, as well as periodicity in peaks.
We hypothesize that these are artifacts caused by the architecture of the \ac{CNN} model, since they are present both in contributions for meaningful sequences (heart \acp{TF}, CTCF), as well as in the negative sequences (with fewer than two corresponding experimental peaks in \ac{ENCODE}).
We thus treat them as a “baseline” signal which we subtract when interpreting the motif contributions in Figure~\ref{fig:contribs_wrt_pos_sub}.
Further investigation of these artifacts - e.g., whether they persist regardless of the employed \ac{CNN} backbone - is an interesting direction for future work.

\section{Enhancer Prediction}
\label{app:enhancer}
We used the scikit-learn Python package~\citep{scikit-learn} to fit the logistic regression models.
The models were optimized for a maximum of $1,000$ iterations per model, using balanced class weights.
For each tissue type, we selected 80\% of samples as training data and the remaining 20\% for evaluation.
We tuned the weights for the $\LNorm_2$ penalty of the logistic regression models with cross-validation on the training subset and evaluated the best-performing model on the test subset.

\subsection{FANTOM5}
\label{sssec:enhancer_fantom}
We used sequences from the 9th and 10th chromosomes of the FANTOM5 dataset~\citep{dalby_2017_556775}. Enhancer sequences in this dataset were identified by an independent experimental assay of ENCODE, therefore it does not contain the exact same biases as the experiments in ENCODE.
We further filtered the samples to match the experiments from \ac{ENCODE}, containing at least 30 positive (enhancer) samples, and obtained the final set of 1459 enhancer sequences from 13 different tissues.



\subsection{VISTA}
\label{sssec:enhancer_vista}
We downloaded $1,940$ human sequences from the Vista Enhancer Browser~\citep{visel2007vista}\footnote{We downloaded the data on 9th August 2021}, which contains $998$ enhancer sequences, and converted them to hg38 coordinates using the liftOver tool~\citep{hinrichs2006ucsc}. These sequences were experimentally tested to have enhancer activity using a reporter assay and therefore, similar to the FANTOM dataset, VISTA enhancers are independent of biases present in ENCODE data.
We selected tissue types with at least 50 positive (enhancer) samples.
Since most sequences are longer than the input length of the \ac{MethodName} model, which has a median length of $1,530$ dinucleotides, we encoded sub-sequences from each sample using a sliding window approach and took the mean of the resulting features as inputs for the logistic regression models.
We report the mean \ac{AUROC} score computed over all tissue types in Table~\ref{tab:vista}.

\begin{table}[t]
\caption{
Results of the enhancer classification task on the VISTA dataset.
For each available tissue type, we train a range of logistic regression models using different features obtained from pretrained \ac{MethodName} models.
We report mean \ac{AUROC} values computed over all tissue types.
}
\label{tab:vista}
\begin{center}
\begin{tabular}{l|lcc}
\hline
       $\coeffCov$ & Combined & Biological & Technical  \\
\hline
      Baseline &  \textbf{0.65} &       - &      - \\
      0 &  \textbf{0.65} &       \textbf{0.65} &      0.64 \\
      0.001 &  0.64 &       \textbf{0.65} &      0.62 \\
      0.01 &  0.64 &       0.64 &      0.62 \\
      0.1 &  0.55 &       0.55 &      0.57\\
\hline
\end{tabular}
\end{center}
\end{table}
\section{Variant Effect Prediction}
\subsection{GTEx}
\label{sssec:gtex}
We retrieved fine-mapped GTEx~\cite{lonsdale2013genotype} \ac{eQTL} variants from the \ac{eQTL} catalog~\cite{kerimov2023eqtl}.
We constructed a positive set of likely causal fine-mapped variants, and a matched negative set, as follows: we excluded all variants that overlap protein-coding genes to limit variants acting through mechanisms other than transcription. 
For other types of transcripts (e.g., lncRNAs), we exclude variants that physically overlap the transcripts they are associated with.
For the positive set, for every tissue, we keep only variants with a posterior inclusion probability (PIP) of $>0.95$.
If variants are also detected in other tissues, we keep them only if they have an average PIP of $>0.5$ across all tissues. 
To sample the negative set, for each positive variant, we identify other variants within a $\pm$5kb window that do not overlap the gene the positive variant is associated with and had PIP $< 0.05$ in all tissues. 
We then select the variant with the most similar allele frequency to the positive variant.
If there are ties based on the allele frequency, we choose the variant that is physically closest to the positive variant within the allele-frequency-matched variants. 
This selection results in a position- and allele-frequency-matched set of $2,304$ negative and $2,339$ positive single nucleotide variants (it can happen that the same negative variant is selected for multiple positive variants across tissues).

We perform variant effect prediction for all variants in this set and the $2,106$ model outputs.
We also select the top 10\%, 5\% and 1\% largest absolute variant effect predictions for each output, and calculate the enrichment (OR) of positive vs negative variants against all other variants. 
We use Fisher's exact test to determine significance.

\subsection{gnomAD}
\label{sssec:gnomad}
We retrieved functionally annotated autosomal genetic variants from Hugging Face \url{https://huggingface.co/datasets/songlab/human_variants} as presented in \cite{benegas2023gpn}.
These variants contain common variants (\ac{MAF} $ > 5\%$) as well as a matched number of rare singleton variants from gnomAD \cite{chen2023genomic}.
We intersect these variants with ENCODE promoter-like cis-regulatory elements \cite{encode2020expanded}.
$44,062$ variants remain after intersection, of which $26,112$ are rare and $17,950$ are common.

We predict variant effects for all variants and $2,106$ model outputs.
For every model output, we calculate variant effect prediction cutoffs at varying thresholds (e.g., the top 0.1\% and 0.01\% most negative/positive values), and calculate odds ratios to quantify the enrichment for rare variants in those extremes vs all other remaining variants. 
We perform Fisher's exact tests to identify significant differences. 

\end{appendices}
\end{document}